\documentclass[a4paper,fleqn,usenatbib]{mnras}

\usepackage{newtxtext,newtxmath,pdflscape}

\usepackage[T1]{fontenc}
\usepackage{ae,aecompl}

\usepackage{graphicx}	
\usepackage{amsmath}	
\usepackage{amssymb}	

\title[Massive stars in transition phases in M33]{On the evolutionary state of massive stars in transition phases in M33\thanks{Based on observations obtained at the Gemini Observatory, under program  GN-2015B-Q-2, which is operated by the Association of Universities for Research in Astronomy, Inc., under a cooperative agreement with the NSF on behalf of the Gemini partnership: the National Science Foundation (United States), the National Research Council (Canada), CONICYT (Chile), Ministerio de Ciencia, Tecnolog\'ia e Innovaci\'on Productiva (Argentina), and Minist\'erio da Ci\^{e}ncia, Tecnologia e Inova\c{c}\~{a}o (Brazil).}}
\author[M. Kourniotis et al.]{M.~Kourniotis$^{1}$,
          M.~Kraus$^{1}$,
          M.~L.~Arias$^{2,3}$, 
          L.~Cidale$^{2,3}$,          
          A.~F.~Torres$^{2,3}$\\
\\
$^{1}$ Astronomick\'y \'ustav, Akademie v\v{e}d \v{C}esk\'e republiky,
Fri\v{c}ova 298, 251\,65 Ond\v{r}ejov, Czech Republic\\
$^{2}$ Instituto de Astrof\'isica de La Plata (CCT La Plata$-$CONICET, UNLP), Paseo del Bosque S/N, La Plata B1900FWA,\\~~~Buenos Aires, Argentina\\ 
$^{3}$ Departamento de Espectroscop\'ia, Facultad de Ciencias Astron\'omicas y Geof\'isicas, Universidad Nacional de La Plata,\\~~~Paseo del Bosque S/N, La Plata B1900FWA, Buenos Aires, Argentina
}

\date{Accepted XXX. Received YYY; in original form ZZZ}

\pubyear{2018}

\begin{document}
\label{firstpage}
\pagerange{\pageref{firstpage}--\pageref{lastpage}}
\maketitle

\begin{abstract}

The advanced stages of several high-mass stars are characterized by episodic mass loss shed during phases of instability. Key for assigning these stars a proper evolutionary state is to assess the composition and geometry of their ejecta alongside the stellar properties. We selected five hot LBV candidates in M33 to refine their classification, investigate their circumstellar environments and explore their evolutionary properties.  Being accessible targets in the near-infrared, we conducted medium-resolution spectroscopy with GNIRS/GEMINI in the $K-$band to investigate their molecular circumstellar environments. Two stars were found to display CO emission, which was modeled to emerge from a circumstellar or circumbinary Keplerian disk/ring. The identification of the carbon isotope $^{13}$C and, for one of the two stars, a significantly low $^{12}$CO/$^{13}$CO ratio, implies an evolved stellar state. As both CO emission stars are highly luminous and hence do not undergo a red supergiant phase, we suggest that stripping processes and equatorial high-density ejecta due to fast rotation are responsible for the enrichment of the stellar surface with processed material from the core. A candidate B[e]SG displays an absorption CO profile, which may be attributed to a jet or stellar pulsations. The featureless infrared spectra of two stars suggest a low-density gas shell or dissipation of the molecule due to the ionizing temperature of the star. We propose spectroscopic monitoring of our targets to evaluate the stability of the CO molecule and assess the time-dependent dynamics of the circumstellar gas structures.
\end{abstract}




\section{Introduction}

Mass loss constitutes the most substantial property that drives massive stars throughout the diverse evolutionary channels \citep{2014ARA&A..52..487S}. The latest stellar models incorporate the updated knowledge on mass loss and provide a way to assign observed stars a current evolutionary state and further predict their fate \citep{2012A&A...537A.146E}. Observational evidence of massive stars exhibiting eruptive events, however, clearly indicate deviations from the conventional and well determined line-driven mass losses that are typically shown during the early phases of the stellar life \citep{2001A&A...369..574V}. Instead, the proximity to the Eddington luminosity limit, the high rotational velocities, encounters with companions, and variability in the outer stellar layers due to pulsations, and energetic shocks are shown to trigger or enhance large amount of expelled gas \citep{2012ARA&A..50..107L,2012A&A...538A..40G}.

Episodic and/or eruptive mass-loss events introduce and characterize the peculiar transient states of Luminous Blue Variables \citep[LBVs,][]{1994PASP..106.1025H}, B[e] supergiant stars \citep[B{[}e{]}SGs,][]{1985A&A...143..421Z}, and Yellow Hypergiants \citep[YHGs,][]{1998A&ARv...8..145D}, which are found to occupy distinct regions in the upper Hertzsprung-Russell (HR) evolutionary diagram. From the evolutionary point of view, B[e]SGs are classified as evolved, post main-sequence hot stars with log$\,L/$L$_{\odot}>4$. They were originally suggested as fast rotators possessing a two-component wind; a fast and low-density polar component and a slow, dense equatorial outflow \citep{1985A&A...143..421Z, 1986A&A...163..119Z, 2005A&A...437..929C}. Observational studies on the geometry of the ejecta confirm the presence of circumstellar disks and rings, which however, are found to be in Keplerian rotation \citep{2012A&A...548A..72C, 2012MNRAS.423..284A, 2016MNRAS.456.1424A, 2017ASPC..508..219K, 2017AJ....154..186K, 2012A&A...543A..77W, 2014AdAst2014E..10D, 2017ASPC..508..213M}. Compared to B[e]SGs, both classes of LBVs and YHGs on average display higher luminosities (log$\,L/$L$_{\odot}\gtrsim 5.4$) with the latter class being constrained by the Humphreys$-$Davidson limit \citep{1994PASP..106.1025H} at log$\,L/$L$_{\odot}\sim 5.8$. \cite{2018MNRAS.478.3138D} recently refined this limit at log$\,L/$L$_{\odot}\sim 5.5$, although their study focused exclusively on the luminous cool population of the Magellanic Clouds. The LBVs in quiescence span an extended region in the HR diagram that is believed to host two separate populations associated with different evolutionary paths; a high-luminosity population formed by LBVs that may still be at their core-hydrogen burning phase, and a low-luminosity population that comprises evolved stars, which have already passed through the red supergiant (RSG) phase and undergo a blueward loop \citep{2013A&A...558A.131G, 2017ApJ...844...40H}. YHGs are also regarded as post-RSG stars \citep{2001ApJ...560..934S} although it is not clear whether they constitute ancestors of LBVs or an end-point state of intermediate-mass massive stars \citep{2014A&A...561A..15C}. The evolutionary link between B[e]SGs and the other two evolved classes has been discussed in many studies \citep[e.g.][]{1996ApJ...468..842S, 1996ApJ...470..597M, 2008A&A...477..193M} but is still under investigation.

From the spectroscopic point of view, these three types of peculiar stars display signatures of gas ejecta indicated by the broad hydrogen emission, which either stands out from H$\alpha$ survey catalogs \citep[e.g.][]{2007AJ....134.2474M} or systematically or occasionally appear during spectroscopic monitoring \citep[e.g.][]{2003ApJ...583..923L}. Low-excitation lines of singly ionized metals, such as \ion{Fe}{II} and [\ion{Fe}{II}] typically make their appearance in the spectra of the three types \citep{1998A&A...340..117L, 2014A&A...561A..15C, 2017ApJ...836...64H}. Emission of [\ion{O}{I}] and [\ion{Ca}{II}] further characterizes the classes of B[e]SGs and YHGs \citep[e.g.][]{2012MNRAS.423..284A,2017ASPC..508..239A,2017ApJ...836...64H}, but is absent or not prominent in LBVs. With respect to their spectral energy distribution (SED), B[e]SGs show a remarkable near-infrared excess arising from warm/hot circumstellar dust \citep{1986A&A...163..119Z,2009AJ....138.1003B, 2010AJ....140..416B}. In contrast, LBVs lack hot dust and typically display free-free emission in the near infrared due to winds. Excess toward longer wavelengths is frequently observed, and implies extended volumes of cool dust formed due to past eruptive events \citep{1986A&A...164..435W,1988ApJ...324.1071M}. YHGs are often enshrouded by warm dust (although not as prominently as B[e]SGs are), and differentiate from the other hot two classes showing a red continuum in the optical that is consistent with A$-$G type stars.  

Characterizing the  stellar properties of the three extreme types is undermined in the sense that, in outbursts, actual photospheres are usually veiled under optically thick material that is formed during severe mass-loss events \citep{1987ApJ...317..760D, 2009ASPC..412...17O}. Alternatively, the properties of the ejected gas e.g. temperature, composition, and density, allow insight on stellar properties such as the intensity of the radiation field and surface abundances. In turn, assessing the internal mixing processes and the geometry of the ejecta could help to infer on the rotational status of the star \citep{2007A&A...464.1029D}. An optimal tracer for the distribution of the expelled gas is found to be the CO molecule observed in the $K-$band \citep{2000A&A...362..158K,2013A&A...549A..28K}. Given its dissociation temperature at 5\,000 K, the CO traces the warm circumstellar gas, which survives the strong radiation field from the hot star \citep{1988ApJ...334..639M}. The study of the $^{13}$C isotope as a product of core-helium burning processes is a step further to shed light on the evolutionary state of the star \citep{2010MNRAS.408L...6L, 2009A&A...494..253K, 2013A&A...549A..28K}. Observations of $^{13}$CO indicate enrichment of the stellar surface via internal mixing or/along with stripping processes induced by enhanced mass losses. Measuring the ratio $^{12}$CO/$^{13}$CO serves as a robust method to evaluate the abundance of processed material \citep[e.g.][]{2009A&A...494..253K,2014ApJ...790...48H,2015AJ....149...13M} with a low ratio  even promoting a post-RSG classification \citep[e.g.][]{2013A&A...558A..17O}.

Nearby galaxies with well established distances offer the opportunity to acquire reliable luminosities, which are essential for discriminating among the various subclasses of B[e] stars \citep{1998A&A...340..117L} and for selecting potential YHGs against their lower mass counterparts and the foreground contaminants \citep{2016ApJ...825...50G}. Moreover and as previously mentioned, luminosity is the main property to tell about different evolutionary origin of LBV stars. Studies in the Magellanic Clouds, M31 and M33 galaxies exploit the accessibility of such targets to undertake optical and infrared studies at resolutions sufficient for assigning a proper classification and identifying molecular enrichment \citep{2013A&A...558A..17O,2014ApJ...780L..10K, 2016A&A...593A.112K, 2014ApJ...790...48H, 2017ApJ...836...64H, 2015MNRAS.447.2459S, 2018A&A...612A.113T}. In this frame, we discuss on the evolutionary state of five hot candidate LBVs in M33 by combining newly presented near-infrared spectroscopy with multi-band photometry. We refine the classification of these stars and suggest scenarios that could be subjects of follow-up studies towards a comprehensive picture of the mass-loss mechanisms. The paper is structured as follows: in Sec. \ref{sample} we describe the input data and discuss on the infrared colors of the targets, in Sec. \ref{infspec} we describe the spectroscopic observations and model the CO first-overtone band heads of two stars showing emission, in Sec. \ref{sed} we model the SEDs of the stars to infer on the stellar and surrounding properties, and in Sec. \ref{disc}, we discuss our results alongside the literature studies. Concluding remarks are given in Sec. \ref{concl}.

\section{Selected sample}
\label{sample}

We selected five hot LBV candidates in M33 from \cite{2007AJ....134.2474M} with the aim to strengthen or refine their classification based on our spectrophotometric exploration. The sample stars are listed in Table \ref{tabphot}. The reference study conducted narrowband imaging centered on the H$\alpha$, [\ion{S}{II}], and [\ion{O}{III}] lines in nine Local Group galaxies and applied photometric criteria to suggest emission-line stars as objects of particular interest. Upon optical spectroscopy, the authors classified as ``hot LBV candidates'' those stars with a spectrum similar to that of known LBVs in M31 and M33, although it was noted to be indistinguishable from that of luminous B[e]SGs. Of our five selected LBV candidates, J013406.63+304147.8 was classified as such already in \cite{1996ApJ...469..629M}, whereas star J013442.14+303216.0 was reported as a questionable LBV due to the lack of forbidden lines in its spectrum. The latter star shows a large $B-V$ color index, which renders it as a potential YHG or an LBV in outburst.

\subsection{Photometric data}
\label{phot}

Photometry from the Local Group Galaxy Survey (LGGS) in the $UBVRI$ was employed from \cite{2016AJ....152...62M}, which contains improved astrometry and revised $U-B$ colors with respect to the initial release of LGGS. For the needs of building the spectral energy distributions of our objects, we complemented the optical observations with data from infrared counterparts using a search radius of $2\arcsec$ from the LGGS coordinates. We retrieved three 2MASS sources, whereas all stars were included in the latest \textit{Spitzer} point-source catalog by \cite{2015ApJS..219...42K}, which contains sources in the IRAC (3.6, 4.5, 5.8, $8\;\mu {\rm{m}}$) and MIPS ($24\;\mu {\rm{m}}$) images in the direction of seven star-forming Local Group galaxies. The resulting photometry in the range $0.37-24\;\mu {\rm{m}}$ is listed in Table \ref{tabphot}. 

Mid-infrared photometry obtained by the Wide-field Infrared Survey Explorer \citep[AllWISE;][]{2014yCat.2328....0C} was additionally explored using a matching radius of $6\arcsec$ equal to the angular resolution of WISE at 3.6 and 4.6 $\mu$m, although not included in our fit study as it could be subject to contamination from nearby sources.
Magnitudes were converted to fluxes using effective wavelengths and zeropoints made available by the SVO Filter Profile Service\footnote{http://svo2.cab.inta-csic.es/theory/fps/}.

\begin{tiny}
\begin{table*}
\caption[]{Photometry and 1-$\sigma$ uncertainties of our five studied stars in M33.}
\label{tabphot} 
\centering 
\begin{tabular}{c|rrrrrrrrrrrrr}
\hline \hline
 LGGS               &  $U$  &  $B$  &  $V$  &  $R$  & $I$ &  $J$ &   $H$  &  $K_{s}$  &   [3.6]  &  [4.5]  &  [5.8]  &  [8.0]  &  [24] \\
\hline
J013248.26+303950.4 & 16.129  & 17.320 & 17.252 & 17.012 & 16.885 &        &        &        & 15.81  & 15.55  & 15.29  & 13.89  & 12.69  \\    
                    &  0.005  & 0.004  &  0.003 &  0.004 &  0.005 &        &        &        &  0.04  &  0.05  &  0.06  &  0.07  &        \\    
J013333.22+303343.4 & 18.305  & 19.296 & 19.397 & 18.439 & 18.888 & 17.06  & 15.92  & 14.79  & 13.25  & 12.69  & 11.94  & 11.30  & 10.08  \\
                    &  0.008  & 0.006  &  0.004 &  0.006 &  0.008 &  0.18  &  0.15  &        &  0.05  &  0.04  &  0.04  &  0.22  &        \\
J013406.63+304147.8 & 15.118  & 16.257 & 16.084 & 15.858 & 15.761 & 15.44  & 15.22  & 14.97  & 14.64  & 14.14  & 13.61  & 10.91  & 8.45   \\
                    &  0.007  & 0.006  &  0.004 &  0.006 &  0.008 &  0.05  &  0.10  &  0.11  &  0.05  &  0.06  &   0.1  &  0.08  &        \\
J013442.14+303216.0 & 18.348  &  18.199& 17.341 & 16.888 & 16.435 & 16.03  & 15.23  & 14.56  & 13.73  & 13.20  &  12.5  & 11.46  & 10.04  \\
                    &  0.009  & 0.007  &  0.005 &  0.007 &  0.009 &  0.08  &        &        &  0.02  &  0.02  &  0.03  &  0.08  &   0.1  \\
J013235.25+303017.6 & 17.093  & 18.052 & 18.007 & 17.871 & 17.706 &        &        &        & 16.35  & 15.89  & 15.38  & 13.88  & 10.65  \\
                    &  0.007  & 0.006  &  0.004 &  0.006 &  0.007 &        &        &        &  0.02  &  0.03  &  0.09  &  0.08  &  0.21  \\
\hline
\end{tabular} 
                                                                                               \end{table*}
\end{tiny}

\subsection{Infrared diagrams}
\label{diag}

\begin{figure*} 
\centering
\includegraphics[width=4.2in]{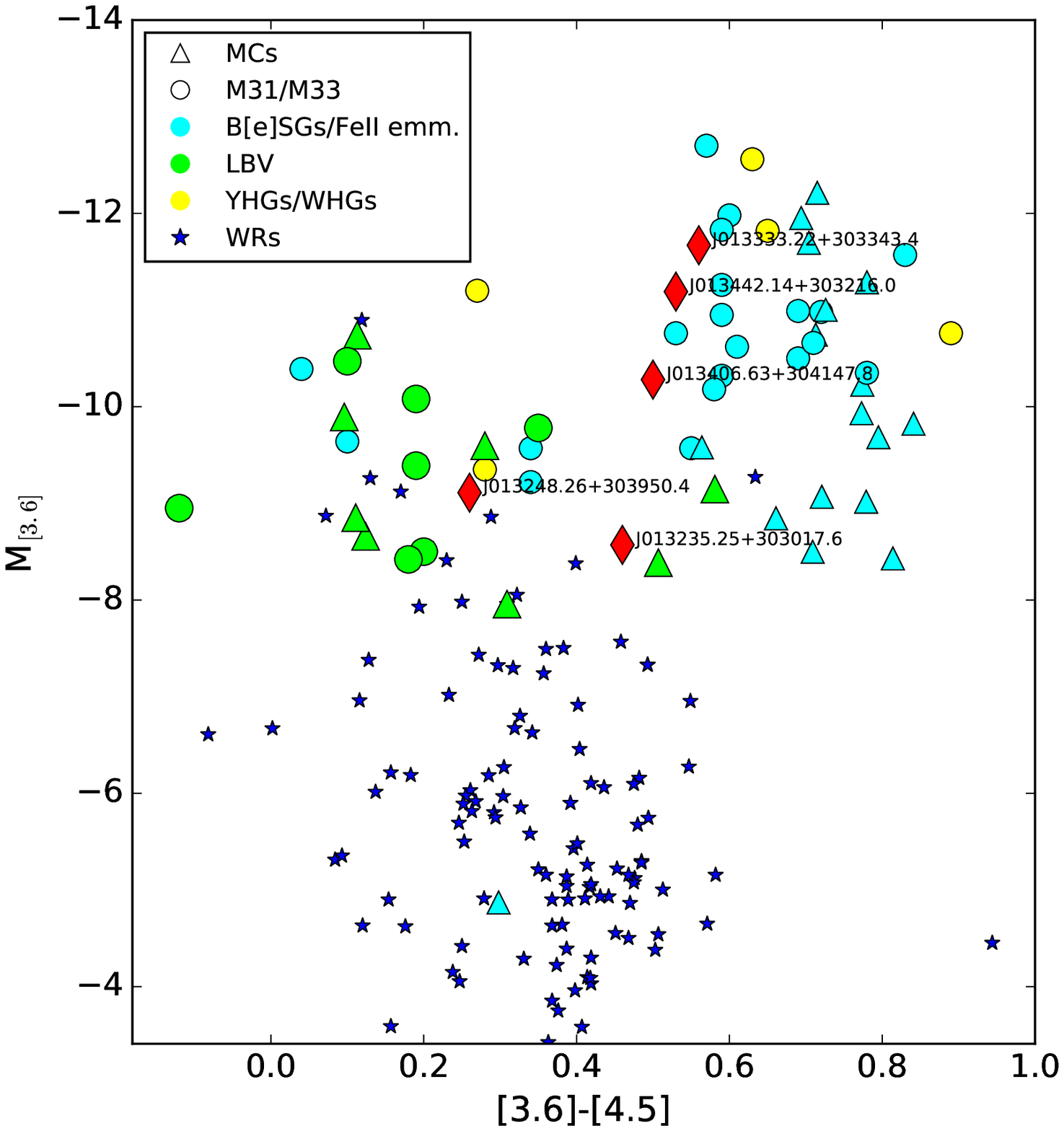}
\caption{M$_{[3.6]}$ vs. [3.6]$-$[4.5] color$-$magnitude diagram for the three extreme stellar types of LBVs (green), B[e]SGs (blue), and YHGs/WHGs (yellow) in the Magellanic Clouds (triangles) and M31/M33 galaxies (circles). The less luminous B[e] star corresponds to LH 85-10 in the LMC (see in the text). Data were taken from \protect\cite{2009AJ....138.1003B, 2010AJ....140..416B} and \protect\cite{2017ApJ...844...40H}. We show the location of our target stars with red diamonds. For comparison, we additionally plot the population of WRs in the Magellanic Clouds.} 
\label{cmd}
\end{figure*}

\begin{figure*} 
\centering
\includegraphics[width=7.2in]{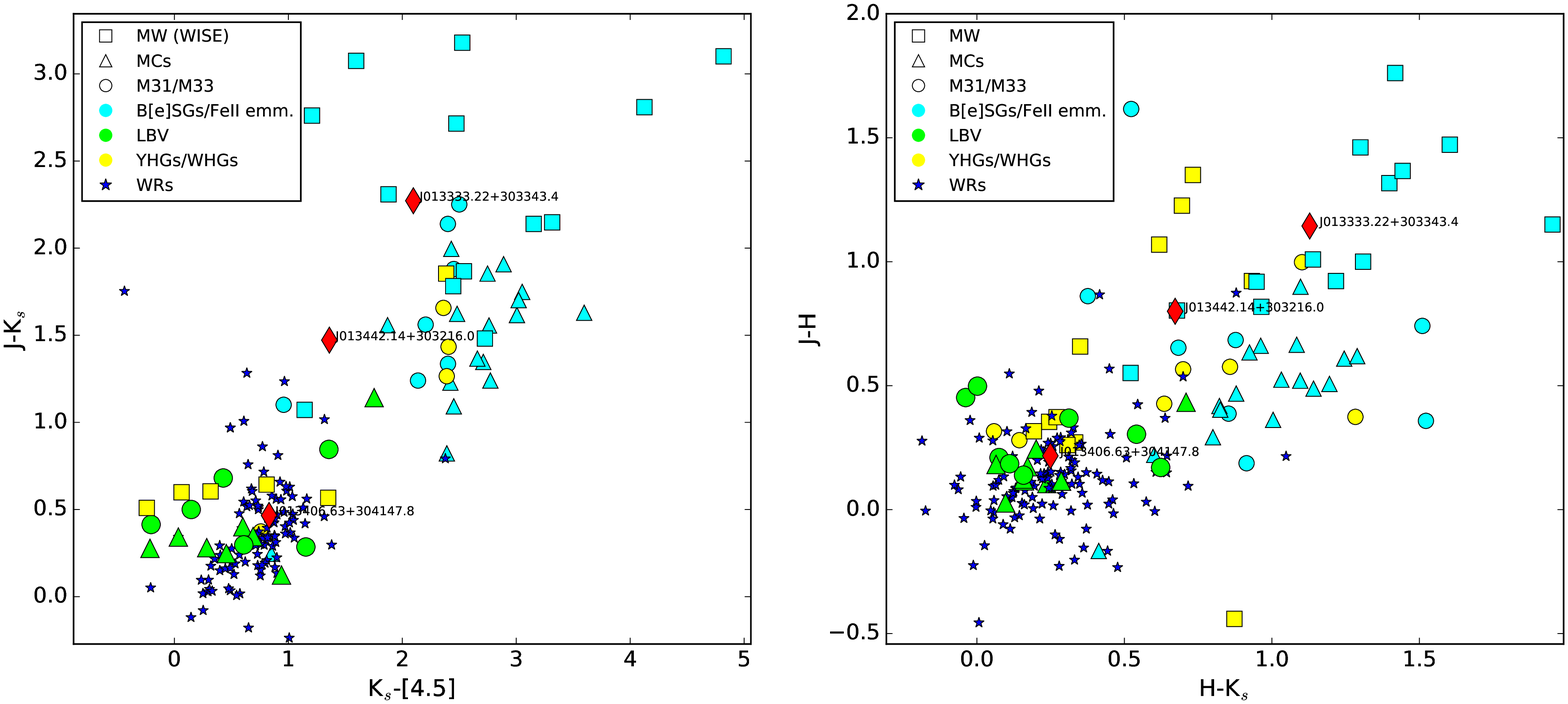}
\caption{J$-$K$_{s}$ vs. K$_{s}-$[4.5] (left) and J$-$H vs. H$-$K$_{s}$ (right) color$-$color diagrams for the stellar types of our interest marked as in Fig. \ref{cmd}. We further show with square symbols the location of the Galactic counterparts \protect\citep{2009A&A...494..253K,1998A&ARv...8..145D}. In absence of \textit{Spitzer} photometry for the latter, we employed WISE W2 magnitudes (at $4.6\;\mu {\rm{m}}$) when available.}
\label{ccd}
\end{figure*}

We built infrared color-magnitude and color-color diagrams for the extreme stellar content in the Local Group. These are displayed in Figs. \ref{cmd} and \ref{ccd}, respectively. The counterparts of evolved stellar classes are shown to stand out from photometric survey catalogs owing to their characteristic infrared colors associated with dust and wind excess \citep[e.g.][]{2009AJ....138.1003B, 2010AJ....140..416B, 2015A&A...584A..33B, 2017A&A...601A..76K}. We show the LBVs and B[e]SGs in the Magellanic Clouds from \cite{2009AJ....138.1003B, 2010AJ....140..416B}, B[e]SGs and YHGs in the Galaxy from \cite{2009A&A...494..253K} and \cite{1998A&ARv...8..145D}, respectively. The B[e]SGs and Warm Hypergiants (WHGs) in M31 and M33 are taken from \cite{2017ApJ...844...40H}. The latter designation stands for dusty luminous post-RSGs with absorption-line spectra typical for A$-$G spectral type stars, which further show emission of forbidden lines \citep{2017ApJ...836...64H}. Although the status of WHGs resembles that of YHGs, they are not found to display the reported spectacular outbursts of the latter, hence we avoid a direct association.

Due to uncertainty in the distance to the Galactic objects, we only show the extragalactic stars in Fig. \ref{cmd}. All B[e]SGs, LBVs, and YHGs/WHGs, are luminous stars with M$_{[3.6]}<-8$ mag. In principle, LBVs separate from B[e]SGs, which occupy a region with $0.6<[3.6]-[4.5]<0.8$ mag indicative of hot dust excess. The outlier B[e]SG LH 85-10 in the LMC, originally classified by \cite{2000AJ....119.2214M}, shows a SED similar to that of classical Be stars \citep{2009AJ....138.1003B}, which we consider to be the most likely classification for the star. We show the population of the luminous and also high mass-loss Wolf-Rayet stars in the Clouds from \cite{2009AJ....138.1003B, 2010AJ....140..416B}, which clearly differentiate from our three studied classes due to their intrinsically bluer colors. The Galactic extreme objects are additionally shown in the color-magnitude diagrams of Fig. \ref{ccd}. In lack of \textit{Spitzer} data, we show the $K_{s}-W2$ colors using the equivalent W2 passband magnitude from WISE at $4.6\;\mu {\rm{m}}$.  The dispersed $K_{s}-W2$ values of the Galactic B[e]SGs are due to the photometric uncertainty of WISE at the saturation limit. Interestingly, they are shown to display particularly red colors, which we attribute to high extinction. Due to their cooler photospheres, YHGs have on average redder 2MASS colors ($J-K_{s}>0.5$ mag) than those of LBVs. The hot dust surrounding B[e]SGs is primarily responsible for their distinct location in the color-color diagrams. On the other hand, the relatively bluer mid-IR colors of LBVs are mainly driven by winds. The majority of the Galactic YHGs show lack of warm dust compared to their M31/M33 counterparts, raising a question on possible effect of metallicity on their mass-loss mechanism. Interestingly, no such correlation is shown for the LBVs in the MCs and M31/M33, which could be attributed to their continuum-driven and thus metallicity-independent mass losses.

Being infrared bright sources, the stars of our sample fit into our scientific objective, with one star being a potential LBV and at least two showing colors characteristic of B[e]SGs. YHGs appear the least constrained class, however, their identification is further supported by their optical colors.

\section{$K-$band spectroscopy}
\label{infspec}

\subsection{Observations and data reduction}
\label{obsspec}

We undertook $K-$band spectroscopy of the five targets with GEMINI/GNIRS. Observations were carried out with the short camera (0.15\arcsec/pix) and the 110.5 l/mm grating, a central wavelength of 2.31 $\mu$m, and a slit width 0.3\arcsec, resulting in spectral resolution $R=5900$. The wavelength coverage spans $2.22-2.40$ $\mu$m. A log of the observations is given in Table \ref{log}. 

\begin{figure} 
\centering
\includegraphics[scale=0.48]{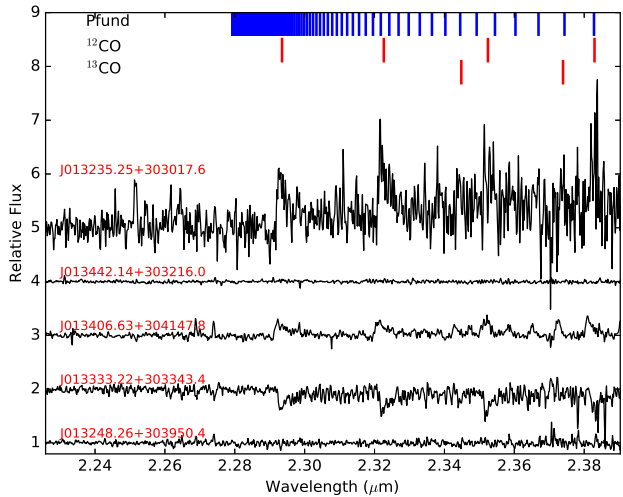}
\caption{$K-$band spectroscopy of our five studied stars in M33. Pfund series and the first$-$overtone bandheads of $^{12}$CO and $^{13}$CO are indicated by vertical ticks.}
\label{figspec}
\end{figure}

Spectra were taken in several ABBA nodding sequences along the slit. The reduction process was carried out using IRAF\footnote{IRAF is distributed by the National Optical Astronomy Observatories, which are operated by the Association of Universities for Research in Astronomy, Inc., under cooperative agreement with the National Science Foundation.} software package tasks. The basic reduction steps were, subtraction of the AB pairs, flatfielding, telluric-correction, and wavelength calibration. A telluric standard star observation was performed immediately prior or after each science target observation. We selected late B and early A-type standard stars due to their few spectral features in the $K-$band. Once the intrinsic stellar lines are removed from the standard star spectrum, we got a telluric template that is used to correct the science target spectrum. The reduced and normalized $K-$band spectra are displayed in Fig. \ref{figspec}. CO emission is found in two stars and absorption in only one. For the two remaining targets, we report a lack of spectral features. 

\begin{table}
\caption[]{Log of near-infrared spectroscopy with GEMINI/GNIRS.}
\label{log}  
\centering
\begin{tabular}{l|cr}
\hline \hline
LGGS  &   Exp. time (s) &   SNR    \\ 
\hline
J013248.26+303950.4  & 4 $\times$ 590 & 30 \\
J013333.22+303343.4  & 4 $\times$ 540 & 20 \\
J013406.63+304147.8  & 4 $\times$ 580 & 25 \\
J013442.14+303216.0  & 4 $\times$ 540 & 45 \\
J013235.25+303017.6  & 4 $\times$ 560 &  5 \\
\hline
\hline
\end{tabular} 
\end{table} 

\subsection{CO band and Pfund line series modeling}
\label{kmodel}

\begin{figure}
\includegraphics[scale=0.48]{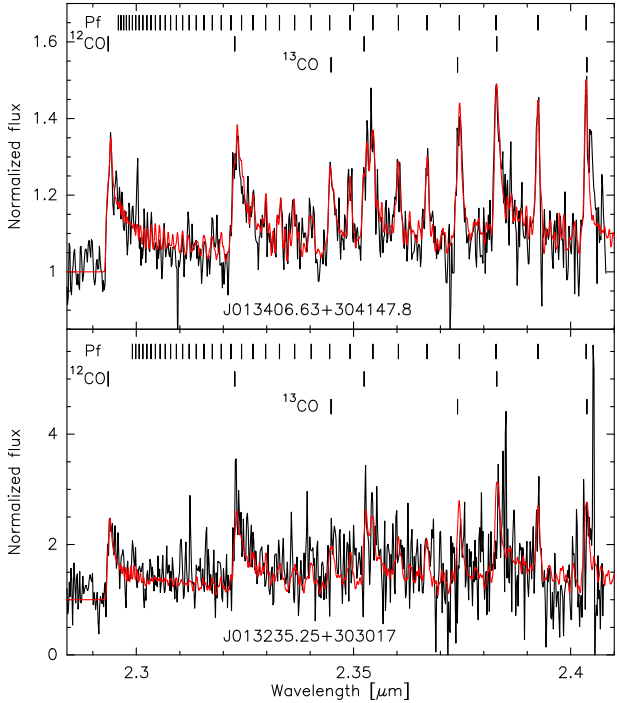}
\caption{Fit to the CO band and Pfund line emission. Ticks mark the positions of individual features.}
\label{CO_fit}
\end{figure}

Two of our objects (J013406.63+304147.8 and J013235.25+303017.6) display CO bands and lines from the hydrogen Pfund series in emission (Fig. \ref{CO_fit}). While the Pfund line emission implies a hot, ionized wind, the molecular gas indicates the co-existence of a cool region, dense enough for efficient molecular condensation, which might indicate the presence of a circumstellar disk. For the modeling we hence follow the same strategy as in previous investigations \citep{2013A&A...558A..17O, 2014ApJ...780L..10K, 2015AJ....149...13M}. 

Starting with the ionized gas region, we use the hydrogen recombination series code of
\citet{2000A&A...362..158K} to model the spectra of the Pfund series. This code follows Menzel's recombination theory \citep{1938ApJ....88..313M} and computes the emission of the Pfund series under the assumption that the lines are optically thin. We fix the electron temperature at $T_{\rm e} = 10\,000$\,K, which is a reasonable value for ionized winds \citep{1991ApJ...377..629L}. Fixing the temperature is no severe restriction, because the line emission is not very sensitive to the temperature. The spectrum of J013235.25+303017 is rather noisy and the Pfund line emission is very weak (see Fig. \ref{CO_fit}), making it difficult to determine the real shape of the hydrogen line profiles. From the Pfund lines in J013406.63+304147.8 we find that a Gaussian line broadening of $\sim$70\,km\,s$^{-1}$ represents the line widths and shapes of the recombination lines reasonably well, and we utilize a Gaussian profile to model the Pfund lines in both objects. High densities within the Pfund line-forming regions lead to pressure ionization, so that the highest levels cannot be populated, and the Pfund series display a sharp cut-off. This level and the corresponding density are included with the other fitting parameters in Table \ref{CO_Pf_results}.

To model the CO bands, we utilize the CO code of \citet{2000A&A...362..158K} suitable to compute the molecular emission from rotating rings and disks in local thermodynamic equilibrium. This code has been advanced by \citet{2009A&A...494..253K} to include the emission from the isotope $^{13}$CO, for which we see clear indication in the spectra of both stars. The shape of the first band head in both objects displays a blue shoulder and a red maximum, although it is less pronounced in J013235.25+303017. Such a shape results from double-peaked profiles of the individual ro-vibrational CO lines and is characteristic for rotating rings or disks \citep{1995Ap&SS.224...25C, 2012A&A...548A..72C, 2000A&A...362..158K, 2013A&A...549A..28K, 2014ApJ...780L..10K, 2015ApJ...800L..20K, 2016A&A...593A.112K, 2015AJ....149...13M, 2018arXiv180700796M}. The sharp onset of the CO band head suggests that the emission formation region is dominated by a single rotation velocity, implying a narrow ring rather than an extended disk. In this case, the free model parameters (CO temperature, column density, and rotation velocity projected to the line-of-sight) are all constant. Constraints for the CO temperature and column densities can be obtained from the strength of the higher band heads and the level of the quasi-continuum between them. In addition,
the strength of the first band head of $^{13}$CO provides the $^{12}$CO/$^{13}$CO ratio and is hence a measure for the stellar surface enrichment in $^{13}$C at the time when the material was released. The parameters for the final fits are summarized in Table \ref{CO_Pf_results}, and the best-fitting models to the observed emission, normalized to the continuum, are depicted in Fig. \ref{CO_fit}.

\begin{table*}
\begin{minipage}{0.94\textwidth}
\caption{$K-$band emission model parameters.}
\label{CO_Pf_results}
\begin{tabular}{lcccccccc}
\hline
Object & \multicolumn{4}{c}{Pfund lines} & \multicolumn{4}{c}{CO bands} \\
           &$v_{\rm gauss,Pf}\sin i$ & $T_{\rm e}$ & n$_{\rm max}$ & $n_{\rm e}$ & $v_{\rm rot,CO}\sin i$  & $T_{\rm CO}$ &  $N_{\rm CO}$  & $^{12}$C/$^{13}$C \\
 & (km\,s$^{-1}$) & (K) &  & (cm$^{-3}$) & (km\,s$^{-1}$) & (K) & ($10^{21}$\,cm$^{-2}$) &  \\
\hline 
J013406.63$+$304147.8   & $70\pm 5$~ & 10\,000 & 59 & $4.8\times 10^{12}$ &  $75\pm 5$~ & $2500\pm 250$~ & $1.0\pm 0.5$ & ~\,$5\pm 1$ \\
J013235.25$+$303017$^a$ & $70\pm 10$ & 10\,000 & 54 & $8.1\times 10^{12}$ &  $50\pm 10$ & $4000\pm 1000$ & $2.0\pm 1.0$ &   $10\pm 2$ \\
   \hline 
\end{tabular}
\smallskip

$^a$ Due to the poor quality of the spectrum, the derived values are only rough estimates.
\end{minipage}
\end{table*}

\section{Spectral energy distributions}
\label{sed}
 
\begin{figure*} 
\centering
\includegraphics[width=\textwidth]{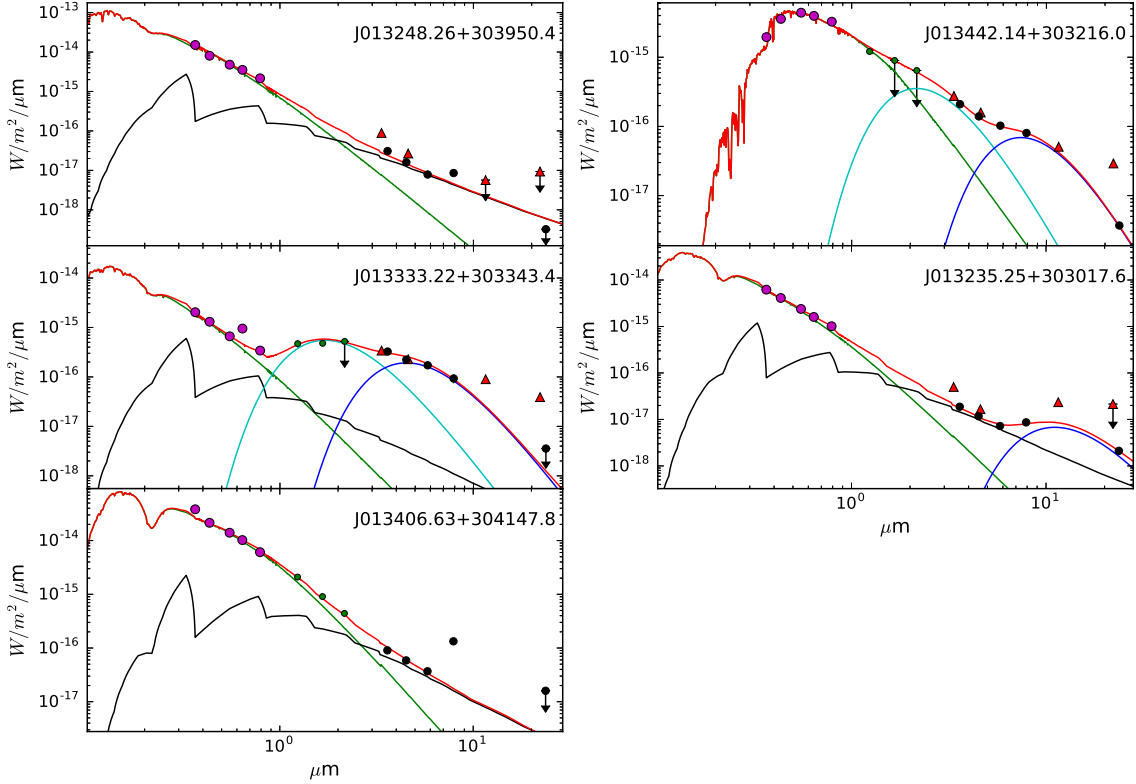}
\caption{Spectral energy distributions for our five luminous stars in M33. We show photometry from LGGS (magenta circles), from 2MASS in $J$, $H$, $K_{s}$ (green circles), IRAC 3.6, 4.5, 5.8, $8\;\mu {\rm{m}}$ and MIPS $24\;\mu {\rm{m}}$ from \textit{Spitzer} (black circles). An arrow indicates an upper limit in the flux. The best-fitting model (red line) is the sum of the reddened, Kurucz photospheric component (green line), a free-free wind model (black line), a hot dust component (cyan line), and a cool dust component (blue line). The parameters corresponding to the fits are provided in Table \ref{tabsed}. Photometry from WISE is additionally plotted (red triangles) but not included in the fits.}
\label{figseds}
\end{figure*}

The SEDs of our target stars display the characteristic near-infrared excess, which is typical of surrounding dust and/or ionizing wind (Fig. \ref{figseds}). Three of the five stars show excess at the longer wavelengths that indicates extended, cool dust envelopes likely formed due to past eruptive activity. 

Modeling the SEDs allows to determine stellar temperatures, retrieve the properties of the dusty environments, and measure bolometric luminosities. The process is thoroughly described in \cite{2017A&A...601A..76K} and is based on a Levenberg-Marquardt minimization scheme for fitting the observed LGGS, 2MASS, and \textit{Spitzer} photometry, setting as free parameters the stellar and dust temperature(s) along with scaling factors as functions of the stellar radius and the emitting surface of the dust, respectively. The dust components were modeled as modified black bodies with a dust emissivity index $\beta=1.5$. Values of the visual extinction were taken from \cite{2014ApJ...790...48H}, delivered either from correcting the reddened colors of known stars located close to our science objects or using the column density maps of neutral hydrogen. Given the large uncertainty in the $A_{V}$ value of the very luminous object J013406.63+304147.8, we exceptionally fixed its temperature at T$_{eff}=30\,000$\,K, in agreement with the late$-$O spectral type of the star from the literature. Extinction laws were taken from \cite{1989ApJ...345..245C}, setting $R_{V}=3.1$. We did not account for additional circumstellar extinction under the hypothesis that the absorbed blue radiation from the surrounding dust is re-emitted in the infrared.

In lack of sophisticated atmospheric models for reproducing atmospheres of the discussed stellar phases, we used a wide range of low-gravity ATLAS9 models from \cite{2011MNRAS.413.1515H}. We chose models at half-solar metallicity similar to that of the LMC, consistent with the location of the stars outside the center of M33 \citep{2005ApJ...635..311U}. The models assume a mixing length to pressure scale height ratio $l/H=0.5$ and a microturbulence of $v_{t}=2$ km~s$^{-1}$ that may, however, vary among supergiants. We stress that an LTE approach would not account for the physical conditions that manifest on the atmospheres of eruptive stars. Whereas non-LTE effects in principle undermine the fit of individual spectral lines, metal blanketing is responsible for veiling part of the ultraviolet continuum. We hence estimated temperatures excluding $U-$band flux from the fit process. We caution that due to the energy conservation, line-blanketing would cause a continuum rise over optical wavelengths, although, observations in the $BVRI$ appear to well reproduce the best-fit model. Exception is J013333.22+303343.4, which shows a discrepant $R-$band measurement that was excluded from the fit and is attributed to a broad H$\alpha$ emission. For the four hot candidates, we complemented the SED fit with a wind spectrum due to free-free radiation and successfully reproduced the observed power-law tail in the infrared. We fixed the electron temperature $T_{\rm e} = 10\,000$\,K. The envelope was reddened according to the aforementioned extinction values and stellar radii were adopted from the fit model. 

Bolometric luminosities were delivered by integrating the unreddened best-fit SED models across the photometry, and regarding that contribution from flux above $24\;\mu {\rm{m}}$ is negligible. The distance to M33 was set to the value $D=964\pm54$ kpc \citep{2006ApJ...652..313B}. Following \cite{2017A&A...601A..76K}, we measured uncertainties using a bootstrap procedure by fitting 1000 sets of photometry and distance uniformly selected within their uncertainties. We further accounted for the half-step error in the temperature of the ATLAS9 models, and quadratically added to the statistical error. We accordingly propagated to the uncertainty of the integrated luminosities. The fit parameters are listed in Table \ref{tabsed}. In Fig. \ref{figseds}, the best-fit photospheric, wind, dust, and total flux models, are plotted over the SEDs of the five studied stars. 

For the two CO emission stars, namely J013406.63+304147.8 and J013235.25+303017.6, we examined the case that a surrounding disk may contribute to the integrated luminosity. Given our lack of knowledge on the inclination, size, and structured composition of the disk, we assumed that both the wind and dust emission originate exclusively from the disk. By excluding these components from the SED of the two stars, the integration of the photospheres yields negligible differences ($<0.01$ dex) from the currently provided luminosities, which are veiled by the error in the distance to M33.

In the following section, we link the inferred physical parameters to the evolutionary phase of the stars in conjunction with the obtained infrared spectroscopy.

\begin{table*}
\caption[]{Parameters of the inferred SED models.}
\label{tabsed}  
{\centering
\begin{tabular}{c|ccccc}
\hline \hline
  LGGS   &  $E(B-V)$   & $T_{\text{eff}}$  &  log$\,L/$L$_{\odot}$ &  Dust & Winds  \\ 
  	 &  (mag) &  (K) &   &  &   \\ 
\hline
J013248.26+303950.4  & 0.4 & 23000 (540) & 6.07 (0.05) & $-$ & yes \\
J013333.22+303343.4  & 0.5 & 29500 (1030) & 5.52 (0.06) & 1300$+$490 K & yes \\
J013406.63+304147.8  & 1.4 (fit) & 30000 (500) & 7.1 (0.06) & cool? & yes \\
J013442.14+303216.0  & 0.5 & 6000 (125) & 5.25 (0.04) & 1030$+$300 K & $-$  \\
J013235.25+303017.6  & 0.8 & 33000 (500) & 6.15 (0.04) & 200 K & yes \\
\hline
\end{tabular}                                                                                   
} 
\end{table*}

\section{Discussion}
\label{disc}

\begin{figure*} 
\centering
\includegraphics[width=7.3in]{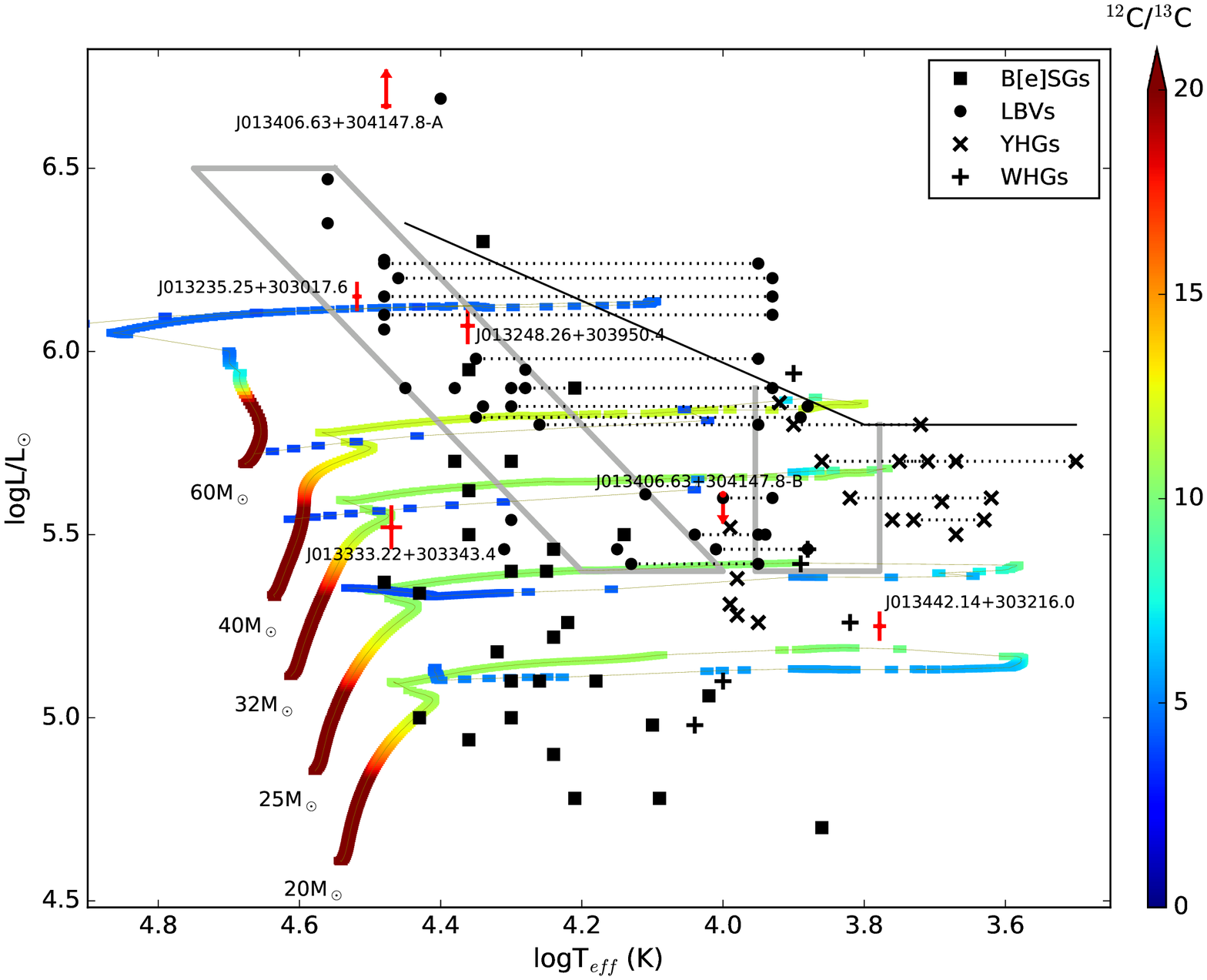}
\caption{Hertzsprung-Russell evolutionary diagram for the LBVs (circles), B[e]SGs (squares), and YHGs/WHGs ('x' symbols) in the Local Group galaxies from the studies of \protect\cite{1998A&ARv...8..145D}, \protect\cite{2004ApJ...615..475S}, \protect\cite{2009A&A...494..253K}, and \protect\cite{2017ApJ...844...40H}. A dotted horizontal line indicates the transition phases for several LBVs and YHGs. We mark with grey line the regions corresponding to the S Dor instability strip and the cooler Yellow Void. The Humphreys-Davidson limit \citep{1994PASP..106.1025H} is shown with black thick line. Evolutionary models at solar metallicity from \protect\cite{2012A&A...537A.146E} are overplotted, assuming rotation at the 40\% of the critical velocity. The grid points of the stellar tracks are color-coded with respect to the $^{12}$C/$^{13}$C ratio. The location and 1-$\sigma$ uncertainties of our discussed stars are displayed with red.}
\label{hrdiag}
\end{figure*}

The key to better understand the complex evolutionary scheme for peculiar stars is to evaluate the observed properties of stars in conjunction with the current theoretical expectations. Populating the HR diagram is essential to refine the regions and limits of instability and update the values of mass-loss rates at critical phases e.g. a cool supergiant phase, which define the posterior evolution of the stars. The incidence of outliers could potentially highlight the role of rotation and binarity. We proceeded to investigate the nature of our five stars over the HR diagram of Fig. \ref{hrdiag}. Overplotted are shown the Geneva evolutionary tracks \citep{2012A&A...537A.146E} for initial masses $20-60$ M$_{\odot}$ at solar metallicity and assuming rotation at the $40\,\%$ of the critical velocity. Each grid/age point of a track is color coded with respect to the expected $^{12}$C/$^{13}$C ratio.

We indicate the location of the reported LBVs, B[e]SGs, and YHGs/WHGs in the Local Group \citep[][and references therein]{1998A&ARv...8..145D, 2004ApJ...615..475S, 2009A&A...494..253K, 2017ApJ...844...40H}. For several LBVs and YHGs, a dotted horizontal line connects the hot quiescent with the cool eruptive state of the star. 
In principle, luminosities are thought to preserve during the instability phases of the star, although \cite{2018A&A...613A..33C} recently reported an increase in the bolometric luminosity of the LBV R40 during the eruption of 2016. The majority of LBVs in quiescence are well confined within the S-Dor instability strip. In their outbursts, counterparts with luminosities above log$\,L/$L$_{\odot}\sim5.8$ are shifted over or even beyond the Humphreys-Davidson limit. The less luminous LBVs are post-main sequence stars that during their eruption may be found within the so called ``Yellow Void''. The latter locus is theoretically predicted to host stars with negative density gradient in part of their extended atmospheres \citep{1995A&A...302..811N, 2001MNRAS.327..452D}. YHGs are believed to bounce against the Yellow Void, exhibiting outbursts that shift them redwards due to the formation of optically thick winds or pseudo-photospheres \citep{1997MNRAS.290L..50D,2009ASPC..412...17O}. The luminosity threshold of log$\,L/$L$_{\odot}\sim5.4$ for a YHG appears to well agree with the observed low-luminosity LBVs, which renders the latter as potential descendants of YHGs. Few WHGs in M31 are found below the luminosity limit (Fig. \ref{hrdiag}), although, and as already stressed in Sec. \ref{diag}, none of these have shown eruptive behavior. Similarly, no outbursts have been reported for the four YHGs below log$\,L/$L$_{\odot}\sim5.4$, three of which are in the LMC and 6 Cas in the Galaxy. On the blue region of the HR diagram, B[e]SGs occupy a wide locus with luminosities up to log$\,L/$L$_{\odot}\sim5.7$, with three stars exceeding further and up to 6.3. Similar to the low-luminosity LBVs, their location on the evolutionary diagram supports their evolved nature, with the exception of a couple of hot stars at log$\,L/$L$_{\odot}\sim5.3$ (Fig. \ref{hrdiag}). Nevertheless, the latter could trace the post-RSG path of stars with M$_{ini}=25$ M$_{\odot}$.

In the following paragraphs, we separately discuss on the evolutionary state of our five target stars.

\begin{itemize}
\item{J013248.26+303950.4 is noted as an iron star by \cite{2012A&A...541A.146C} showing signatures of warm circumstellar dust due to the excess at $8\;\mu {\rm{m}}$. Our SED fit, however, indicates infrared photometry consistent with free-free emission from winds, although data from WISE may not exclude the presence of cool dust (Fig. \ref{figseds}). Given the constraint at $24\;\mu {\rm{m}}$, the excess at $8\;\mu {\rm{m}}$ is likely due to PAH emission instead of a dust component. The optical spectroscopy by \cite{2017ApJ...836...64H} indicates emission of \ion{He}{I} and H$\alpha$ lines, which are characteristic of the supergiant class. The presence of \ion{Fe}{II} and [\ion{Fe}{II}] lines renders the star as counterpart of our studied stellar types. The inferred luminosity of log$\,L/$L$_{\odot}\sim6.1$ that places the star within the S Dor strip along with the absence of [\ion{O}{I}] \citep{2017ApJ...836...64H} favor a classification of an LBV in quiescence instead of a B[e]SG or YHG.

The featureless $K-$band spectrum of the star is similar to that of LBVs \citep{2013A&A...558A..17O} implying conditions that prevent the appearance of CO lines. The photospheric Pfund lines are filled with emission originating in the ionizing wind, which is responsible for the photodissociation of the CO molecule. As also seen in Fig. \ref{figseds}, the wind dominates the near-infrared continuum of the star. \cite{1988ApJ...324.1071M} stated that CO association could occur at large distances around hot luminous stars, when the strong stellar winds compress a remnant shell ejecta to sufficiently high densities. The data from \textit{Spitzer} however, do not point out a dust envelope that could result from a previous mass-loss event, which makes less likely that our $K-$band spectrum captures a variable CO profile.}

\item{Of our less luminous objects, J013333.22+303343.4, stands out due to its significant infrared excess requiring two different dust components; a hot dust envelope of 1\,300 K located close to the star, and a cool dust volume of 500 K extending further. The star shows an evident $R-$band excess due to a broad H$\alpha$ emission. Free-free emission from winds provides fit to the $I-$band measurement, which deviates from the photospheric component. Integrating the SED model yields log$\,L/$L$_{\odot}=5.5$, placing the star at the end of its core-hydrogen burning phase or at the hot phase of a blueward evolutionary loop (Fig. \ref{hrdiag}). The spectroscopic study of \cite{2017ApJ...836...64H} reports emission of \ion{He}{I}, \ion{Fe}{II}, [\ion{Fe}{II}], [\ion{O}{I}], which is typically shown in the spectrum of B[e]SGs. Nebular lines of [\ion{N}{II}] are reported by \cite{2014ApJ...790...48H}. 

The infrared spectrum of J013333.22+303343.4 is one of the few of the candidate B[e]SGs showing the CO band heads in absorption. The carbon isotope $^{13}$C is present, however, due to the low S/N ratio we can not resolve a $^{12}$CO/$^{13}$CO lower than 10. Consequently, we can not conclude whether the star is departing from the main sequence or is an evolved post-RSG star. Whereas a stable CO emission in B[e]SGs is associated with a dense equatorial disk/ring, absorption implies that the molecular gas interferes between the observer and the star. A jet oriented along the line of sight could stand as a plausible explanation. \cite{2016A&A...585A..81M} discovered a jet emerging from the projected center of the Galactic MWC 137, a B[e]SG that is also known to host a gaseous circumstellar disk. The current $K-$band spectrum of J013333.22+303343.4 does not support the presence of a circumstellar disk that could fuel the jet as may happen in the case of MWC 137, although it may not exclude filling-in CO emission. Another interpretation is that possible pulsational activity of the star could modulate the spectroscopic profile of the surrounding gas, which is expelled in the form of a shell ejecta. Expansion thus cooling of the gas behind the pressure front of a pulsationally-driven shockwave, could give rise to a periodical CO absorption. A $K-$band monitoring throughout the possible pulsational cycle is needed to resolve such mechanism. Accordingly, a profile variability in the \ion{He}{I} line of the B[e]SGs LHA 120-S 73 and LHA 120-S 35 was suggested to be due to stellar pulsations \citep{2016A&A...593A.112K,2018A&A...612A.113T}.

It is tempting to assume that CO absorption may not emerge from the gaseous surroundings but rather, it has photospheric origin. \cite{2014MNRAS.443..947L} reported a CO absorption spectrum for the Galactic B[e] stars MWC 623 and AS 381, which rises from a cool (K$-$type) supergiant companion. Both B[e] stars are less luminous than J013333.22+303343.4 (log$\,L/$L$_{\odot}<5$), which makes this scenario less likely for our case but, at the same time, more appealing. We examined the case that the SED model incorporates the spectrum of a cool secondary of temperature $2\,000-5\,000$ K, which is necessary to give rise to the molecular CO lines. The coolest reported RSG in M33 has a temperature of 4\,300 K \citep{2012ApJ...750...97D} and can drop as low as 2\,700 K in the LMC \citep{2012ApJ...749..177N}. Due to the degeneracy between the hot dust temperature and that of the cool stellar body, we can not set tight constraints on the latter. The marginal value of 2\,700 K yields the highest possible luminosity for the companion, namely log$\,L/$L$_{\odot}=4.5$ (corresponding to a star with M$_{ini}\sim9$ M$_{\odot}$), and the addition of more dust shells and/or of a wind model would further lower this value. As seen from Fig. \ref{binary2}, a putative such companion contributes only $\sim10\,\%$ to the total $K-$band flux, which conflicts with the strong CO band heads that reach half the continuum flux (Fig. \ref{figspec}). We hence conclude that a cool companion is less likely to explain the current case of molecular absorption.

\begin{figure} 
\centering
\includegraphics[scale=0.47]{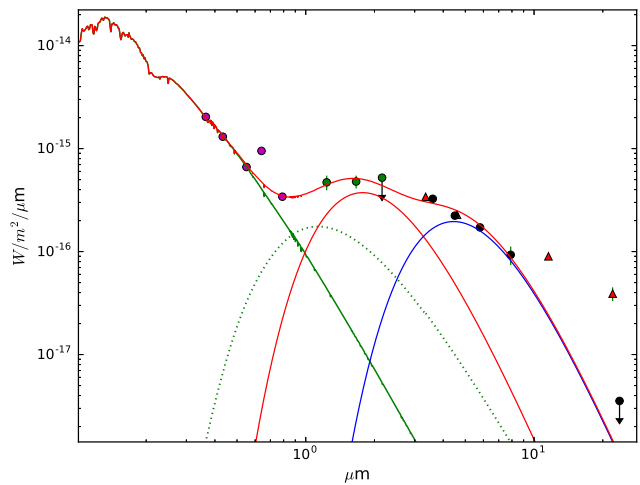}
\caption{Composite SED fit model of J013333.22+303343.4. The dotted green line corresponds to a black body of 2\,700 K to account for an extreme cool supergiant secondary.}
\label{binary2}
\end{figure}
}

\item{J013406.63+304147.8 or B416 is a remarkably luminous star in M33. Our SED fit indicates emission from ionizing winds and a steep increase at $8\;\mu {\rm{m}}$. By inspecting the IRAC maps, the target is found embedded within extended nebulosity at $8\;\mu {\rm{m}}$ although being well resolved at the shorter wavelengths. We hence attribute the $8\;\mu {\rm{m}}$ offset to a strong PAH contamination. A considerable difference of 1.7 mag between the PSF and aperture photometry at $8\;\mu {\rm{m}}$ is further reported by \cite{2015ApJS..219...42K}. Optical spectroscopy shows an \ion{H}{II} region-like spectrum with emission of Balmer lines, [\ion{O}{III}], and [\ion{S}{II}] \citep{1996AJ....112.1450C}. Emission of \ion{He}{I}, \ion{Fe}{II}, and [\ion{Fe}{II}] was reported by \cite{1996ApJ...469..629M}, who classified the star as an LBV candidate due to its spectroscopic and color similarity with Var C. Nevertheless, no significant variability that is characteristic of an erupting LBV has been yet reported \citep{2000MNRAS.311..698S, 2012A&A...541A.146C}. Alternatively, the classification as a massive B[e]SG \citep{2005A&A...437..217F} is questionable due to the lack of hot dust that is typically associated with the class.

The star is surrounded by an expanding ring-like nebula, which is suggested by \cite{2005A&A...437..217F} to be the remnant of main-sequence or pre-LBV bubble. \cite{2000MNRAS.311..698S} precluded the nebula from being a circumstellar shell but instead, the star should be located at the center of an \ion{H}{II} region. The same study assigned B416 a periodic variability of 8.25 days and further reported long-term variations in the continuum of the star. The case that periodicity originates in eclipsing configurations of a binary was discussed, however, due to the lack of spectroscopic signatures from the companion it was regarded as rather unlikely. On the other hand, by measuring the variability in the radial velocities of \ion{He}{I} and H$\alpha$ lines, \cite{2004BaltA..13..156S} detected a period of 16.13 days, almost double the previously reported value, and concluded that B416 is a close interacting binary with a mass ratio of 0.4.

The $K-$band spectrum of the star reveals CO overtone emission implying the presence of a dense shielding structure, which prevents the molecule from dissociation due to the strong radiation from the central star. We favor the case of a circumbinary ring over that of circumstellar disk. Based on its high luminosity beyond the S Dor strip, it is reasonable to consider J013406.63+304147.8 as a pair of stars rather than a single star \citep{2014ApJ...790...48H}. We proceeded to evaluate the contribution from two photospheres to the overall fit of the SED. We considered the case were both stellar components contribute equally to the visual flux in an attempt to provide a lower limit for the luminosity of the primary. We fixed the temperature of the primary at 30\,000 K consistent to a late$-$O spectral type \citep{2017ApJ...836...64H}, and repeated the process discussed in Section \ref{sed}, with the addition of a secondary photospheric component. Visual extinction was set as a free parameter. The best fit to the photometry was then achieved with a 10\,000 K secondary and A$_{V}=0.9$ mag (Fig. \ref{binary}). At this marginal case, luminosities were found to be log$\,L_{1}/$L$_{\odot}=6.67$ and log$\,L_{2}/$L$_{\odot}=5.61$ for the primary and secondary, respectively. By increasing the contribution of the primary to the total flux, extinction was increased up to the value of 1.4 mag, which corresponds to the single star case (Table \ref{tabsed}). Comparing to evolutionary models in Fig. \ref{hrdiag}, we infer that the hidden secondary reaches a maximum M$_{ini}=32$ M$_{\odot}$. These parameters raise a critical discrepancy concerning the coevality of the components. Based on the theoretical predictions, even if found in a pre-RSG state the secondary should be older than 6.7 Myr. This value is unrealistic for a notably luminous primary, which surpasses the luminosity of a 120 M$_{\odot}$ star the lifetime of which does not exceed 3 Myr. In the frame of stellar encounters, B416 could have been a binary with components of similar mass that underwent a rapid mass exchange. At its current state, the system consists of a very massive and overluminous for its mass star, with the donor star being significantly fainter, even below its detection limit. This scenario not only justifies the luminosity of a star being otherwise too high for a single star, but also supports the ejection of gas from the system. Non-conservative Roche lobe overflow of the donor star would result in a circumbinary outflow, which consists of gas that is enhanced in $^{13}$CO \citep[e.g.][]{2013A&A...549A..28K}. The measured low $^{12}$CO/$^{13}$CO value could serve as an indicator of this process.

\begin{figure} 
\centering
\includegraphics[scale=0.48]{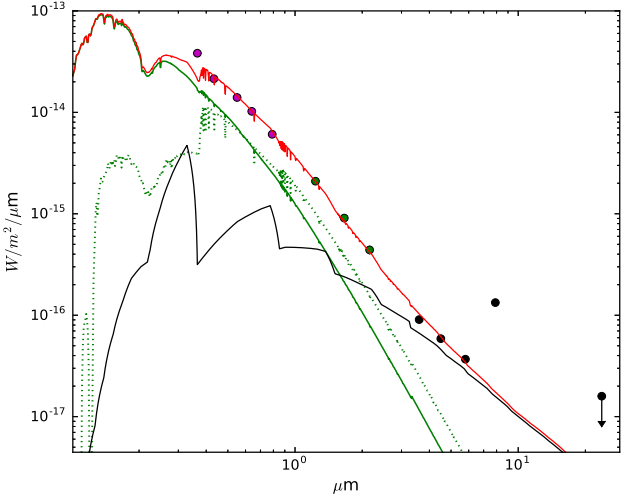}
\caption{Composite SED fit model of J013406.63+304147.8. The dotted green line represents a secondary stellar component of 10\,000 K, which is scaled to match the optical flux of the primary component (solid green line).}
\label{binary}
\end{figure}
}
\item{Denoted as a hot LBV candidate by \cite{2007AJ....134.2474M}, J013442.14+303216.0 is the coolest of our five studied stars. Its temperature of 6\,000 K prevents classification as an LBV, even in eruptive phase. Based on the optical spectroscopy of \cite{2017A&A...601A..76K}, the star shows narrow H$\alpha$ and nebular [\ion{S}{II}] lines. In the same study, the authors concluded that its relatively low luminosity along with the narrow hydrogen profile are not consistent to the properties of a YHG. Our $K-$band spectroscopy lacks evidence of a shielding circumstellar structure as opposed to the emission CO profile of the two YHGs in the LMC sample of \cite{2013A&A...558A..17O}. For non-disk ejecta, the star should have only recently departed from a cool mass-loss phase in order to prevent the physical expansion of the CO shell \citep{1988ApJ...334..639M}. We conclude that star J013442.14+303216.0 is either a post-RSG surrounded by diluted or low-mass molecular gas, or a star that is still evolving redwards the HR diagram. 

From Fig. \ref{hrdiag}, J013442.14+303216.0 shares the same evolutionary track with several reported B[e]SGs. \cite{2013A&A...558A..17O} stated that CO is not detected for B[e]SGs below log$\,L/$L$_{\odot}=5$ as a consequence of their low mass-loss rates and thus, their insufficiently low density circumstellar environments. Nevertheless, three stars above that limit showed absence of CO implying either diversity in the geometry of the circumstellar structures or even a questionable stellar classification. Assigning B[e]SGs with absence of CO a post-RSG status, they may constitute descendants of stars like J013442.14+303216.0. As the latter is not expected to undergo the instability phase of a YHG in order to further shed large amounts of mass, the star is expected to cross the HR diagram bluewards retaining a featureless $K-$band spectrum in the region of the CO bands.}

\item{J013235.25+303017.6 is classified as an iron star by \cite{2012A&A...541A.146C}. Expanding gas shell gives rise to a \ion{He}{I} P Cygni profile, and emission of [\ion{N}{II}] indicates circumstellar nebulosity \citep{2014ApJ...790...48H}. A lack of diagnostic lines for B[e]SGs, such as [\ion{Fe}{II}] and [\ion{O}{I}], is reported by \cite{2017ApJ...836...64H}. The data from \textit{Spitzer} trace emission from ionizing winds and further indicate the presence of cool dust at longer wavelengths. Along with these features, our inferred high luminosity of log$\,L/$L$_{\odot}=6.15$ render the star a potential LBV in quiescence.

J013235.25+303017.6 is the second of our observed objects that displays the CO molecule in emission. Modeling of the $K-$band spectrum favors the presence of a dense circumstellar disk, although the low S/N ratio of the data may undermine a robust conclusion. Theoretical models at solar metallicity predict that a 60 M$_{\odot}$ star has already reached a low ratio $^{12}$CO/$^{13}$CO ($<5$) during its post-main sequence lifetime (Fig. \ref{hrdiag}). Our modeled value is found higher than the predicted, which implies that the surrounding molecular gas could have been structured already at the early stellar phases. On the other hand, half-solar metallicity models predict that a $^{12}$CO/$^{13}$CO $\sim10$ is expected when stars have already encountered the Humphreys$-$Davidson limit. A past instability phase may justify ejection of material that formed the currently observed cool dust shell, although it is questionable whether the duration of such transition phase is long enough for the star to establish a large CO envelope. CO emission is not anticipated in the spectrum of LBVs, however, exceptional cases such as HR Car have been reported, showing a variable CO profile that switches from emission to absorption within a short interval \citep{1997ASPC..120...20M}. Emission of CO arising from ejected shells (e.g. from LBV eruptions) must be short-lived due to the fast dilution of the gas to densities below the value that is critical for collisional excitation \citep{1988ApJ...334..639M}. Long-lasting CO emission may point out another mechanism that preserves gas to sufficient densities, such as the disks surrounding Oe stars \citep{1974ApJ...193..113C, 2007IBVS.5773....1R}. Intrinsic rotation close to the critical velocity would establish a high-density equatorial outflow already at the early stellar lifetime \citep{2017arXiv171202908K} and, at the same time, enhance internal mixing processes for the efficient enrichment of the surface with core-burning products \citep{2007A&A...464.1029D}. On departure from the main sequence, the angular momentum loss would slow down the star and consequently halt the replenishment of the disk. Posterior to this point, further processed material would be dissociated due to strong radiation field. This may provide explanation for the discrepancy between the measured $^{12}$CO/$^{13}$CO presumably originating from an early-formed disk and the expected low value from the models. Interestingly, the location of J013235.25+303017.6 on the evolutionary diagram is similar to that of the peculiar Oe star LHA 120-S 124 in the LMC, which also shows a prominent and stable CO overtone emission \citep{2013A&A...558A..17O}. The latter star was characterized as an active LBV \citep{1998A&A...332..857V}, although \cite{1987A&A...181..293S} classified it as a late$-$O supergiant with a circumstellar disk, as inferred from the observed double-peak \ion{Fe}{II} emission.}
\end{itemize}

\section{Conclusions}
\label{concl}

We undertook $K-$band spectroscopy of five luminous stars in M33 characterized by high mass-loss rates, to gain insight into their gaseous surroundings and to better understand the advanced stages in the lifetime of extreme massive stars. The objects are bright targets in the infrared with M$_{[3.6]}<-8$ mag, occupying regions where B[e]SGs, LBVs, and YHGs reside. By modeling the SED of our stars with photospheric, wind and dust components, we estimated effective temperatures and derived bolometric luminosities. With accurate distance measurements in hand, these extragalactic targets well confine their location on the evolutionary diagram thus allowing a direct comparison to theoretical models. 

We report two stars, J013248.26+303950.4 and J013442.14+303216.0, showing absence of the CO molecule. Of these, the spectrum of the luminous LBV candidate J013248.26+303950.4 indicates prevention of molecular formation due to the hot temperature of the star. For the less luminous warm supergiant, insufficient gas densities or a non-recent RSG phase could justify observations. The hot B[e]SG candidate J013333.22+303343.4 displays a remarkable absorption CO profile, which is attributed to gas interfering between the observer and the star e.g. a bipolar stream. Alternatively, stellar pulsations may be responsible for a variant CO profile depending on the compression or expansion phase of the circumstellar gas. The remaining two stars in our sample show emission in the CO bandheads. Of these, we suggest the hotter, J013235.25+303017.6, as an Oe star where emission emerges from an early-formed circumstellar disk. The most luminous star of our sample, J013406.63+304147.8, leaves room for a faint cool companion, which could support both the striking luminosity of the main star owing to mass transfer and the formation of a circumbinary disk/ring, which consists of processed material ejected from the system.

The current near-infrared data can serve as milestone to follow-up studies, which will assess the evolutionary picture of the five extreme stars and their analogues. Spectroscopic monitoring of the objects is needed to evaluate the stability of the CO molecule. A variable profile could be investigated in the frame of pulsation cycles, interaction of the molecular circumstellar gas with shockwaves, or the physical expansion, and thus dilution of the gas volume. In addition, optical high-resolution spectroscopy on [\ion{Ca}{II}] and [\ion{O}{I}] lines will allow insight into the kinematics and disk/ring structures \citep[e.g.][]{2012MNRAS.423..284A,2016MNRAS.456.1424A,2017ASPC..508..239A,2016A&A...593A.112K,2017ASPC..508..213M,2018A&A...612A.113T}. Profile variability as a consequence of orbital motion within a binary could reveal the dominant role of stellar encounters to the rotating status of the discussed stars, the chemical enrichment through deposit or stripping processes and the formation of circumbinary rings. Along with time-series photometry obtained by long-term monitoring surveys such as Pan-STARRS \citep{2016arXiv161205560C} and Gaia \citep{2016A&A...595A...1G}, future studies could associate phases of enhanced mass loss to stellar and environmental modulations.

\section*{Acknowledgements}

We thank the anonymous referee for the valuable comments, which helped us to improve our manuscript. M.~Kraus acknowledges financial support from GA\,\v{C}R under grant number 17-02337S. The Astronomical Institute Ond\v{r}ejov is supported by the project RVO:67985815. LC thanks financial support from the Agencia de Promoci\'on Cient\'ifica y Tecnol\'ogica (Pr\'estamo BID PICT 2016/1971), CONICET (PIP 0177), and the Universidad Nacional de La Plata (Programa de Incentivos G11/137), Argentina. This research has made use of NASA's Astrophysics Data System Bibliographic Services and the VizieR catalogue access tool, CDS, Strasbourg, France. 





\bibliographystyle{mnras}
\bibliography{kourniotis18_final}




\bsp	
\label{lastpage}
\end{document}